\documentclass[pre,twocolumn,aps,showpacs]{revtex4}
\newcommand{\bec}[1]{\mbox{\boldmath $ #1$}}
\usepackage{graphicx}
\begin{document}
\bigskip
\bigskip
\title{Shear-current effect in a turbulent convection
with a large-scale shear}
\author{Igor Rogachevskii}
\email{gary@bgu.ac.il} \homepage{http://www.bgu.ac.il/~gary}
\author{Nathan Kleeorin}
\email{nat@menix.bgu.ac.il} \affiliation{Department of Mechanical
Engineering, The Ben-Gurion University of the Negev, POB 653,
Beer-Sheva 84105, Israel}
\date{\today}

\begin{abstract}
The shear-current effect in a nonrotating homogeneous turbulent
convection with a large-scale constant shear is studied. The
large-scale velocity shear causes anisotropy of turbulent
convection, which produces the mean electromotive force $\bec{\cal
E}^{(W)} \propto {\bf W} {\bf \times} {\bf J}$ and the mean electric
current along the original mean magnetic field, where ${\bf W}$ is
the background mean vorticity due to the shear and ${\bf J}$ is the
mean electric current. This results in a large-scale dynamo even in
a nonrotating and nonhelical homogeneous sheared turbulent
convection, whereby the $\alpha$ effect vanishes. It is found that
turbulent convection promotes the shear-current dynamo instability,
i.e., the heat flux causes positive contribution to the
shear-current effect. However, there is no dynamo action due to the
shear-current effect for small hydrodynamic and magnetic Reynolds
numbers even in a turbulent convection, if the spatial scaling for
the turbulent correlation time is $\tau(k) \propto k^{-2}$, where
$k$ is the small-scale wave number.
\end{abstract}

\maketitle

\section{Introduction}

It is generally believed that the large-scale magnetic fields of the
Sun, solar type stars and galaxies are originated by a dynamo
process, i.e., due to a joint action of small-scale helical
turbulent motions (the $\alpha$ effect) and large-scale differential
rotation (see \cite{M78,P79,KR80,ZRS83,RSS88,S89,RS92,O03,BS05}). It
has been recently recognized \cite{RK03,RK04,B05,BHK05,RKL06,RKCL06}
that in a sheared nonhelical and nonrotating homogeneous turbulence
whereby the $\alpha$ effect vanishes, the mean-field dynamo is
possible due to shear-current effect.

The mechanism of the shear-current dynamo is as follows. The
deformations of the original nonuniform magnetic field lines are
caused by upward and downward turbulent eddies. In a sheared
turbulence the inhomogeneity of the original mean magnetic field
breaks a symmetry between the influence of the upward and downward
turbulent eddies on the mean magnetic field. This creates the mean
electric current ${\bf J}$ along the mean magnetic field ${\bf B}$
and produces the shear-current dynamo. In particular, the
large-scale velocity shear creates anisotropy of turbulence. This
produces the mean electromotive force $\bec{\cal E}^{(W)} \propto
{\bf W} {\bf \times} {\bf J}$, where ${\bf W}$ is the background
mean vorticity due to the shear. Joint effects of the mean
electromotive force $\bec{\cal E}^{(W)}$ and stretching of the mean
magnetic field due to the large-scale shear motions cause the
mean-field dynamo instability.

A sheared turbulence is a universal feature in astrophysics. The
shear-current effect might be an origin for the large-scale magnetic
fields in colliding protogalactic clouds and in merging protostellar
clouds \cite{RKCL06}. This effect might be also important in
accretion discs where the mean velocity shear comes together with
rotation, so that both the shear-current effect and the $\alpha$
effect might operate. Since the shear-current effect is not quenched
(see \cite{RK04,RKA06}) contrary to the quenching of the nonlinear
$\alpha$ effect, the shear-current effect might be the only
surviving effect, and it can explain the origin of large-scale
magnetic fields in astrophysical plasmas with large-scale sheared
motions.

The shear-current effect is a fundamental phenomenon which should be
studied in different situations, e.g., in different types of
turbulence. The goal of the present study is to investigate the
shear-current effect in a nonrotating homogeneous turbulent
convection with a large-scale constant velocity shear. Note also
that in many astrophysical applications turbulent convection plays
an important role, e.g., in the convective zones of the Sun and
solar type stars. We have shown that turbulent convection promotes
the shear-current dynamo instability. In particular, the heat flux
causes positive contribution to the shear-current effect. However,
the shear-current dynamo is impossible for small hydrodynamic or
magnetic Reynolds numbers even in a turbulent convection, if the
spatial scaling for the turbulent correlation time is $\tau(k)
\propto k^{-2}$, where $k$ is the small-scale wave number.

This paper is organized as follows. In section~II we formulate the
governing equations, the assumptions and the procedure of the
derivations. In section~III we study properties of the shear-current
effect in a sheared turbulent convection and discuss the
shear-current dynamo. In section~IV we draw concluding remarks and
perform comparison of the theoretical predictions with the direct
numerical simulations. Finally, in Appendix~A we perform a detailed
derivation of the shear-current effect in a turbulent convection.

\section{The governing equations}

In order to study the shear-current effect in a turbulent convection
we use a procedure which is similar to that applied in
\cite{RK04,RK06}. In particular, we employ a mean field approach
whereby the pressure, entropy, velocity and magnetic fields are
separated into the mean and fluctuating parts, where the fluctuating
parts have zero mean values. To determine the effect of shear on a
turbulent convection we use equations for fluctuations of velocity,
magnetic field and entropy
\begin{eqnarray}
{\partial {\bf u} \over \partial t} &=& - ({\bf U} {\bf \cdot}
\bec{\nabla}) {\bf u} - ({\bf u} {\bf \cdot} \bec{\nabla}) {\bf U} -
\bec{\nabla} \biggl({p \over \rho_{0}}\biggr) - {\bf g} \, s
\nonumber\\
& & + {1 \over \rho_0} \, [({\bf b} {\bf \cdot} \bec{\nabla}) {\bf
B}+ ({\bf B} {\bf \cdot} \bec{\nabla}){\bf b}] + {\bf u}^{N} \;,
\label{B1} \\
{\partial {\bf b} \over \partial t} &=& ({\bf B} {\bf \cdot}
\bec{\nabla}){\bf u} - ({\bf u} {\bf \cdot} \bec{\nabla}) {\bf B} +
({\bf b} {\bf \cdot} \bec{\nabla}){\bf U} - ({\bf U} {\bf \cdot}
\bec{\nabla}) {\bf b} + {\bf b}^N \,,
\nonumber\\
\label{B2} \\
{\partial s \over \partial t} &=& - {N^{2} \over g} ({\bf u} \cdot
{\bf e}) - ({\bf U} \cdot \bec{\nabla}) s + s^{N} \;, \label{B3}
\end{eqnarray}
where ${\bf B}$, $\, {\bf U}$ and $S$ are the mean magnetic field,
the mean velocity field and the mean entropy, ${\bf u}$, $\, {\bf
b}$ and $s$ are fluctuations of velocity, magnetic field and
entropy, $\rho_0$ is the fluid density, $N^{2} = - {\bf g} {\bf
\cdot} \bec{\nabla} S$, and $\, {\bf g}$ is the acceleration of
gravity, ${\bf e}$ is the unit vector directed opposite to ${\bf
g}$, the magnetic permeability of the fluid is included in the
definition of the magnetic field, $p$ are the fluctuations of total
(hydrodynamic and magnetic) pressure, ${\bf v}^{N}$, $\, {\bf b}^{N}
$ and $\, s^{N}$ are the nonlinear terms which include the molecular
viscous and diffusion terms. Equations~(\ref{B1})-(\ref{B3}) for
fluctuations of fluid velocity, entropy and magnetic field are
written in the Boussinesq approximation. We consider the hydrostatic
nearly isentropic basic reference state. The turbulent convection is
regarded as a small deviation from a well-mixed adiabatic reference
state.

Using Eqs.~(\ref{B1})-(\ref{B3}) written in a Fourier space we
derive equations for the instantaneous two-point second-order
correlation functions of the velocity fluctuations $\langle u_i \,
u_j\rangle$, the magnetic fluctuations $\langle b_i \, b_j \rangle$,
the entropy fluctuations $\langle s \, s\rangle$, the cross-helicity
tensor $\langle b_i \, u_j \rangle$, the turbulent heat flux
$\langle s \, u_i \rangle$ and $\langle s \, b_i \rangle$. The
equations for these correlation functions are given by
Eqs.~(\ref{B6})-(\ref{B11}) in Appendix A. We split the tensor of
magnetic fluctuations into nonhelical, $h_{ij} = \langle b_i \, b_j
\rangle$, and helical, $h_{ij}^{(H)},$ parts. The helical part
$h_{ij}^{(H)}$ depends on the magnetic helicity, and it is
determined by the dynamic equation which follows from the magnetic
helicity conservation arguments (see, e.g.,
\cite{KR82,KRR94,GD94,KR99,KMRS2000,BF00,VC01,BB02}, and a review
\cite{BS05}).

The second-moment equations include the first-order spatial
differential operators $\hat{\cal N}$  applied to the third-order
moments $M^{(III)}$. A problem arises how to close the system, i.e.,
how to express the set of the third-order terms $\hat{\cal N}
M^{(III)}$ through the lower moments $M^{(II)}$ (see, e.g.,
\cite{O70,MY75,Mc90}). We use the spectral $\tau$ approximation
which postulates that the deviations of the third-moment terms,
$\hat{\cal N} M^{(III)}({\bf k})$, from the contributions to these
terms afforded by the background turbulent convection, $\hat{\cal N}
M^{(III,0)}({\bf k})$, are expressed through the similar deviations
of the second moments, $M^{(II)}({\bf k}) - M^{(II,0)}({\bf k})$:
\begin{eqnarray}
\hat{\cal N} M^{(III)}({\bf k}) &-& \hat{\cal N} M^{(III,0)}({\bf
k})
\nonumber\\
&=& - {1 \over \tau(k)} \, [M^{(II)}({\bf k}) - M^{(II,0)}({\bf k})]
\;, \label{AAC3}
\end{eqnarray}
(see \cite{O70,PFL76,KRR90,KMR96,RK04}), where $\tau(k)$ is the
scale-dependent relaxation time, which can be identified with the
correlation time of the turbulent velocity field. The quantities
with the superscript $(0)$ correspond to the background  shear-free
turbulent convection with a zero mean magnetic field. We apply the
spectral $\tau$ approximation only for the nonhelical part $h_{ij}$
of the tensor of magnetic fluctuations. Note that a justification of
the $\tau$ approximation for different situations has been performed
in numerical simulations and analytical studies in
\cite{BS05,BF02,FB02,BK04,BSM05,SSB07}.

We assume that the characteristic time of variation of the mean
magnetic field ${\bf B}$ is substantially larger than the
correlation time $\tau(k)$ for all turbulence scales. This allows us
to get a stationary solution for the equations for the second-order
moments, $M^{(II)}$. We split all second-order correlation
functions, $M^{(II)}$, into symmetric $h_{ij}^{(s)} = [h_{ij}({\bf
k}) + h_{ij}(-{\bf k})] / 2$ and antisymmetric $h_{ij}^{(a)} =
[h_{ij}({\bf k}) - h_{ij}(-{\bf k})] / 2$ parts with respect to the
wave vector ${\bf k}$. For the integration in ${\bf k}$-space we
have to specify a model for the background shear-free turbulent
convection (i.e., a turbulent convection with ${\bf B} = 0)$. The
background turbulent convection is maintained by an imposed vertical
heat flux $F^\ast_z = \langle s \, u_z \rangle$ with div$\, {\bf
F}^\ast=0$ at a low boundary of convective region. We used the
following model for the homogeneous background turbulent convection:
\begin{eqnarray}
\langle u_i \, u_j \rangle^{(0)}({\bf k}) &=& \langle {\bf u}^2
\rangle \, P_{ij}(k) \, W(k) \;,
\label{B15} \\
\langle b_i \, b_j \rangle^{(0)}({\bf k}) &=& \langle {\bf b}^2
\rangle \, P_{ij}(k) \, W(k) \;,
\label{B16} \\
\langle s \, u_i \rangle^{(0)}_{i}({\bf k}) &=& 3 \, \langle s \,
u_z \rangle \, e_{m} \, P_{im}(k) \, W(k) \;, \label{B17}
\end{eqnarray}
where $P_{ij}(k) = \delta_{ij} - k_i k_j / k^2$, $\, \delta_{ij}$ is
the Kronecker tensor, $\, W(k) = E(k) / 8 \pi k^{2} ,$ the energy
spectrum is $E(k) = (q-1) \, (k / k_{0})^{-q} ,$ $\, k_{0} = 1 /
l_{0}$ and the length $\, l_{0}$ is the maximum scale of turbulent
motions. The turbulent correlation time is $\tau(k) = C \, \tau_0 \,
(k / k_{0})^{-\mu}$, where the coefficient $C=(q-1+\mu)/(q-1)$. This
value of the coefficient $C$ corresponds to the standard form of the
turbulent diffusion coefficient in the isotropic case, i.e.,
$\eta_{_{T}} = \int \tau(k) \, [\langle {\bf u}^2 \rangle \, E(k)]
\, dk = \tau_0 \, \langle {\bf u}^2 \rangle /3$. Here the time
$\tau_0 = l_{0} / \sqrt{\langle {\bf u}^2 \rangle}$ and
$\sqrt{\langle {\bf u}^2 \rangle}$ is the characteristic turbulent
velocity in the scale $l_{0}$. For the Kolmogorov's type background
turbulence (i.e., for a turbulence with a constant energy flux over
the spectrum), the energy spectrum $E(k) \propto - d\tau / dk$, the
exponent $\mu=q-1$ and the coefficient $C=2$. In the case of a
turbulence with a scale-independent correlation time, the exponent
$\mu=0$ and the coefficient $C=1$. Motions in the background
turbulent convection are assumed to be non-helical. Using the
solution of the derived second-moment equations, we determine the
contributions to the mean electromotive force, ${\cal
E}_{i}^{\sigma} = \varepsilon_{imn} \, \int \langle b_n \, u_m
\rangle_{\bf k} \,{\rm d} {\bf k}$, caused by the sheared turbulence
(see Appendix A), where $\varepsilon_{ijk}$ is the fully
antisymmetric Levi-Civita tensor. This procedure allows us to
determine the contributions to the shear-current effect caused by
the sheared turbulent convection.

\section{The shear-current dynamo}

We consider a homogeneous turbulent convection with a constant mean
velocity shear, ${\bf U} = (0, Sx, 0)$ and $ {\bf W} = (0,0,S)$. We
consider a most simple form of the mean magnetic field,  ${\bf B} =
(B_x(z), B_y(z), 0)$. The contributions to the mean electromotive
force caused by the sheared turbulence, are ${\cal E}_{i}^{\sigma} =
b_{ijk}^{\sigma} \, \nabla_k \, B_{j}$, where the tensor
$b_{ijk}^{\sigma} = b_{ijk}^{u} + b_{ijk}^{F}$ is given by
\begin{eqnarray}
b_{ijk}^{u} &=& l_0^2 \, {I_2  \over 30} \, \varepsilon_{ikn} \,
[Q_0 \, \nabla_n \, U_{j} + 2 \, Q_1 \, (\partial U)_{nj}] \,,
\label{B18}\\
b_{ijk}^{F} &=&  a_\ast \, l_0^2 \, {I_3  \over 140} \, \biggl[Q_3
\, \varepsilon_{ikm} \, e_{mn} \, \nabla_n \, U_{j} + [Q_2 \,
\varepsilon_{ikn}
\nonumber\\
& & + Q_4 \, (\varepsilon_{ikm} \, e_{mn} + \varepsilon_{inm} \,
e_{mk})] \, (\partial U)_{nj} \biggr] \, .\label{B19}
\end{eqnarray}
Equations~(\ref{B18}) and~(\ref{B19}) are derived in Appendix A.
Here $(\partial U)_{ij} = (\nabla_i U_{j} + \nabla_j U_{i}) / 2$ and
the coefficients $Q_n$ are $Q_0 = (3 - 2 \, \mu) - \epsilon \, (5 +
2 \, \mu)$, $\, Q_1 = \epsilon \, (7 + 6 \, \mu)  - 1$, $\, Q_2 =
\mu + 2$, $\, Q_3 = 18 - 19 \mu$, $\, Q_4 = \mu - 6$, the parameter
$\epsilon = E_m / E_v$, $\,\, E_m$ and $E_v$ are the magnetic and
kinetic energies per unit mass in the background turbulent
convection, and
\begin{eqnarray*}
I_2 &=& \int \tau^2(k) \, E(k) \, dk = {(q-1+\mu)^2 \over (q-1+2 \mu
) \, (q-1)} \;,
\\
I_3 &=& \int \tau^3(k) \, E(k) \, dk = {(q-1+\mu)^3 \over (q-1+3 \mu
) \, (q-1)^2} \; .
\end{eqnarray*}
For the Kolmogorov's type turbulence, the exponent $\mu=q-1$ and the
parameters $I_2=4/3$ and $I_3=2$. In the case of a turbulence with a
scale-independent correlation time, the exponent $\mu=0$ and the
parameters $I_2=I_3=1$. The tensor $b_{ijk}^{F}$ in Eq.~(\ref{B19})
describes the contributions of the heat flux to the shear-current
effect, while tensor $b_{ijk}^{u}$ determines the non-convective
contributions (which are independent of the heat flux) to the
shear-current effect. In Eqs.~(\ref{B18})-(\ref{B19}) we have taken
into account only the terms which contribute to the shear-current
effect. In particular, we have taken into account that $B_y \gg B_x$
(see Eq.~(\ref{M6}) below) and considered a weak mean velocity shear
${\bf U} = (0, Sx, 0)$, where $S \tau_0 \ll 1$. The convective
contribution to the dynamo instability due to the shear-current
effect depends on the parameter $a_\ast = 2 g \tau_0 F^\ast_z /
\langle {\bf u}^2 \rangle$ which is determined by the budget
equation for the total energy. The parameter $a_\ast$ is given by
\begin{eqnarray}
a_\ast^{-1} = 1 + { \nu_{_{T}} (\nabla \, U)^2 + \eta_{_{T}} (\nabla
\, B)^2 /\rho_0 \over g \, \langle s \, u_z \rangle} \;,
 \label{AAC1}
\end{eqnarray}
where $\nu_{_{T}}$ is the turbulent viscosity and $\eta_{_{T}}$ is
the coefficient of turbulent magnetic diffusion.

Therefore, in the kinematic approximation the mean magnetic field is
determined by
\begin{eqnarray}
{\partial B_x \over \partial t} &=& - \sigma_{_{B}} \, S \, l_0^2 \,
B''_y + \eta_{_{T}} \, B''_x  \;,
\label{E2}\\
{\partial B_y \over \partial t} &=& S \, B_x + \eta_{_{T}} \, B''_y
\;, \label{E3}
\end{eqnarray}
where $B''_i = \partial^2 B_i / \partial z^2 $. Here we neglect
small contributions to the coefficient of turbulent magnetic
diffusion caused by the shear motions because $S \tau_0 \ll 1$. The
dimensionless parameter $\sigma_{_{B}}$ describes the shear-current
effect. Straightforward calculations using
Eqs.~(\ref{B18})-(\ref{B19}) yield the parameter $\sigma_{_{B}} =
\sigma^u_{_{B}} + \sigma^F_{_{B}}$, where
\begin{eqnarray}
\sigma^u_{_{B}} &=& {I_2  \over 30} \, (Q_0 + Q_1) \;,
\label{SB20}\\
\sigma^F_{_{B}} &=&  a_\ast \, {I_3 \over 280} \, \biggl[Q_2 - Q_4 +
2 \, (Q_3 + Q_4) \, \sin^2 \, \phi\biggr] \;,
\nonumber\\
\label{SB21}
\end{eqnarray}
$\phi$ is the angle between the unit vector ${\bf e}$ and the
background vorticity ${\bf W}$ due to the large-scale shear.
Equations~(\ref{SB20}) and~(\ref{SB21}) yield the following final
expressions for the parameter $\sigma_{_{B}}$:
\begin{eqnarray}
\sigma_{_{B}} &=& {I_2  \over 15} \, \biggl[1 - \mu + \epsilon \, (1
+ 2 \, \mu) + {a_\ast \, 3 \, I_3 \over 7 \, I_2} \, \big[2
\nonumber \\
&& + 3 \, (2 - 3 \mu) \, \sin^2 \, \phi \big]\biggr] \;, \label{E1}
\end{eqnarray}
where the terms $\propto a_\ast$ in Eq.~(\ref{E1}) describe the
contribution of the turbulent convection to the shear-current
effect. Equations~(\ref{E2}) and~(\ref{E3}) determine the
shear-current dynamo instability. In particular, the first term $
\propto S B_x $ in the right hand side of Eq.~(\ref{E3}) determines
the stretching of the magnetic field $B_x$ by the shear motions and
produces the field $B_y$. On the other hand, the interaction of the
non-uniform magnetic field $B_y$ with the background vorticity ${\bf
W}$ produces the electric current along the field $B_y$. This effect
is determined by the first term in the right hand side of
Eq.~(\ref{E2}) and causes the generation of the magnetic field
component $B_x$. The growth rate of the mean magnetic field due to
the shear-current dynamo instability is given by
\begin{eqnarray}
\gamma = S \, l_0 \, \sqrt{\sigma_{_{B}}} \, K_z - \eta_{_{T}} \,
K_z^2  \;,
\label{E65}
\end{eqnarray}
where $K_z$ is the large-scale wave number. The necessary condition
for the dynamo instability is $\sigma_{_{B}} > 0$.

The shear-current dynamo instability depends on the spatial scaling
of the correlation time $\tau(k) \propto k^{-\mu}$ of the turbulent
velocity field, where $k$ is the small-scale wave number. In the
absence of turbulent convection, the terms $\propto a_\ast$ in
Eq.~(\ref{E1}) vanish, and the shear-current dynamo in a
non-convective turbulence with $\epsilon=0$ occurs for $\mu < 1$.
For the Kolmogorov's type turbulence, the exponent $\mu=2/3$ and
Eq.~(\ref{E1}) reads
\begin{eqnarray}
\sigma_{_{B}} = {4  \over 135} \, \biggl[1 + 7 \, \epsilon  + {6
\over 7} \, a_\ast \biggr] \; . \label{E1A}
\end{eqnarray}
In this case the parameter $\sigma_{_{B}}$ is independent of the
angle $\phi$ between the unit vector ${\bf e}$ and the background
vorticity ${\bf W}$. For a turbulent convection with a
scale-independent correlation time, the exponent $\mu=0$ and the
parameter $\sigma_{_{B}}$ is given by
\begin{eqnarray}
\sigma_{_{B}} = {1  \over 15} \, \biggl[1 + \epsilon \, + {9 \over
7} \, a_\ast \, (1 + 3 \sin^2 \, \phi) \biggr] \; . \label{E1B}
\end{eqnarray}
In these cases the shear-current dynamo instability causes the
generation of the large-scale magnetic field. It follows from
Eqs.~(\ref{E1})-(\ref{E1B}) that turbulent convection promotes the
shear-current dynamo instability. In particular, the heat flux
causes positive contribution to the shear-current effect when $2 + 3
\, (2 - 3 \mu) \, \sin^2 \, \phi > 0$.

However, for small hydrodynamic and magnetic Reynolds numbers, the
turbulent correlation time is of the order of $\tau(k) \propto
1/(\nu k^2)$ or $\tau(k) \propto 1/ (\eta k^2)$ depending on the
magnetic Prandtl number, i.e., $\tau(k) \propto k^{-2}$. In this
case $\mu=2$, and the parameter $\sigma_{_{B}} < 0$ even in a
turbulent convection with $\epsilon=0$. This implies that for small
hydrodynamic and magnetic Reynolds numbers there is no dynamo action
due to the shear-current effect. This result is in agreement with
the recent studies \cite{RAS06,RUK06} performed in the framework of
the second order correlation approximation (SOCA) for sheared
non-convective flows. This approximation is valid only for small
hydrodynamic Reynolds numbers. Even in a highly conductivity limit
(large magnetic Reynolds numbers), SOCA can be valid only for small
Strouhal numbers, while for large hydrodynamic Reynolds numbers (for
a developed turbulence), the Strouhal number is 1.

Note that the standard approach (i.e., SOCA) cannot describe the
situation in principle. The reason is that the shear-current dynamo
requires a finite correlation time of turbulent velocity field, so
the delta-correlated version of SOCA fails. The application of the
path integral approach for the study of the shear-current dynamo
also requires a finite correlation time of turbulent velocity field.
The shear-current dynamo is a phenomenon that results from the
interaction of the energy-containing-scale of turbulence with
large-scale shear, and the constraint is that the hydrodynamic and
magnetic Reynolds numbers should be not small at least. Therefore,
the SOCA-based approaches do not work properly to describe the
shear-current dynamo. Probably, the hydrodynamic and magnetic
Reynolds numbers can be of the order of unity and there is no need
for a developed inertial range in order to maintain the
shear-current dynamo.

In order to determine the threshold required for the excitation of
the shear-current dynamo instability, we consider the solution of
Eqs.~(\ref{E2}) and~(\ref{E3}) with the following boundary
conditions: ${\bf B}(t,|z|=L) = 0$ for a layer of the thickness $2L$
in the $z$ direction. The solution for the mean magnetic field is
determined by
\begin{eqnarray}
B_y(t,z) &=& B_0 \, \exp(\gamma \, t) \, \cos (K_z z + \varphi) \;,
\label{M5} \\
B_x(t,z) &=& l_0 \, K_z \, \sqrt{\sigma_{_{B}}} \, B_y(t,z) \; .
\label{M6}
\end{eqnarray}
For the symmetric mode the angle $\varphi =\pi \, n$ and the
large-scale wave number $K_z=(\pi / 2) (2m + 1)\, L^{-1}$, where $n,
m = 0, 1, 2, ... \,$. For this mode the mean magnetic field is
symmetric relative to the middle plane $z=0$. Let us introduce the
dynamo number $D= (l_0 \, S_\ast / L)^2 \, \sigma_{_{B}}$, where
parameter $S_\ast = S \, L^2 / \eta_{_{T}}$ is the dimensionless
shear number. For the symmetric mode the mean magnetic field is
generated due to the shear-current effect when the dynamo number $D
> D_{\rm cr} = (\pi^2/4) (2m + 1)^2$. For the antisymmetric mode the
angle $\varphi =(\pi / 2) \, (2n+1)$ with $n = 0, 1, 2, ...$, the
wave number $K_z=\pi \, m \, L^{-1}$ and the magnetic field is
generated when the dynamo number $D > D_{\rm cr} = \pi^2 \, m^2$,
where $m = 1, 2, 3, ... \,$. The maximum growth rate of the mean
magnetic field in the shear-current dynamo instability, $
\gamma_{\rm max} = S^2 \, l_0^2 \, \sigma_{_{B}} / 4 \eta_{_{T}}$,
is attained at $ K_z = S \, l_0 \, \sqrt{\sigma_{_{B}}} / 2
\eta_{_{T}}$. Therefore, the characteristic scale of the mean
magnetic field variations $L_B = 2 \pi  / K_z = 4 \, u_0 / (S \,
\sqrt{\sigma_{_{B}}}) \, $. For the shear-current dynamo, the ratio
of the field components $B_x / B_y$ is small (see Eq.~(\ref{M6})).
Remarkably, in the $\alpha \Omega$ dynamo, the poloidal component of
the mean magnetic field is much smaller than the toroidal field.

\section{Discussion}

In the present study we investigate the shear-current effect in a
nonrotating homogeneous turbulent convection with a large-scale
constant velocity shear. We show that the condition for the
shear-current dynamo is independent of the exponent of the energy
spectrum of turbulent convection, but it depends on the scaling
exponent $\mu$ of the turbulent correlation time $\tau(k) \propto
k^{-\mu}$, where $k$ is the small-scale wave number. We discuss
three cases in details: (i) the Kolmogorov's type turbulence with
the exponent $\mu=2/3$; (ii) a turbulent convection with a
scale-independent correlation time $(\mu=0)$; (iii) a turbulent
convection with small hydrodynamic and magnetic Reynolds numbers
with the scaling $\tau(k) \propto k^{-2}$. We have found that
turbulent convection promotes the shear-current dynamo instability.
In particular, the heat flux causes positive contribution to the
shear-current instability. However, the shear-current dynamo does
not occur for small hydrodynamic and magnetic Reynolds numbers even
in a turbulent convection, if the spatial scaling for the turbulent
correlation time is $\tau(k) \propto k^{-2}$.

For simplicity we consider weak linear velocity shear, ${\bf U} =
(0, Sx, 0)$, where the parameter $S \tau_0 \ll 1$. The main effect
of the weak linear velocity shear on turbulent convection is a
generation of additional anisotropic velocity fluctuations. We
consider turbulent convection in the region which is far from the
boundaries, because the constant linear velocity shear cannot exist
near the boundaries whereby the boundary layers form. The generation
of the magnetic field in a nonlinear velocity shear depends on
boundary conditions and requires numerical study. Turbulent
convection can be inhomogeneous in this case.

The main goal of this paper is to study an effect of the heat flux
on the shear-current dynamo instability in a most simple model of
turbulent convection with a linear shear. The shear-current dynamo
acts also in inhomogeneous turbulent convection. However, in
inhomogeneous turbulence with a large-scale constant velocity shear
the kinetic helicity and the $\alpha$ effect do not vanish (see
\cite{RK03,RK04,RAS06}). In this case the shear-current dynamo acts
together with the $\alpha$-shear dynamo which is similar to the
$\alpha \Omega$ dynamo. The joint action of the shear-current and
the $\alpha$-shear dynamos have been recently discussed in
\cite{RKA06,P07,BS05B}. The shear-current effect does not quenched
(see \cite{RK04,RKA06}) contrary to the quenching of the nonlinear
$\alpha$ effect, the turbulent magnetic diffusion, the effective
drift velocity, etc. Therefore, the shear-current effect might be
the only surviving effect, and it can explain the origin of
large-scale magnetic fields in sheared astrophysical turbulence.

The shear-current dynamo instability is saturated by the nonlinear
effects. The nonlinear mean-field dynamo due to a shear-current
effect in a nonhelical homogeneous turbulence with a mean velocity
shear has been investigated recently in \cite{RKL06} (see also
\cite{BS05B}), whereby the transport of magnetic helicity as a
dynamical nonlinearity has been taken into account. The magnetic
helicity flux strongly affects the saturated level of the mean
magnetic field in the nonlinear stage of the dynamo action. In
particular, numerical solutions \cite{RKL06} of the nonlinear
mean-field dynamo equations which take into account the
shear-current effect, show that if the divergence of the magnetic
helicity flux is not small, the saturated level of the mean magnetic
field is of the order of the equipartition field determined by the
turbulent kinetic energy. These results are in a good agreement with
direct numerical simulations \cite{B05,BHK05}, whereby the
generation of the large-scale magnetic field in a nonhelical
turbulence with an imposed mean velocity shear has been
investigated.

In the direct numerical simulations \cite{B05,BHK05} the
non-convective turbulence is driven by a forcing that consists of
eigenfunctions of the curl operator with the wavenumbers $4.5 < k_f
< 5.5$ and of large-scale component with wavenumber $k_1=1$. The
forcing produces the mean flow $U= U_0 \, \cos \, (k_1 \, x) \, \cos
\, (k_1 \, x)$. The numerical resolution in these simulations is
$128 \times 512 \times 128$ meshpoints, and the parameters used in
these simulations are as following: the magnetic Reynolds number
${\rm Rm} = u_{\rm rms} / (\eta \, k_f) =80$, the magnetic Prandtl
number ${\rm Pr}_m = \nu/ \eta = 1$ and $U_0 /u_{\rm rms} =5$. The
growth rate of the mean magnetic field is about $\gamma \, \tau_0
\approx 2 \times 10^{-2}$. This allows us to estimate the parameter
$\sigma_{_{B}}$ characterizing the shear-current effect,
$\sigma_{_{B}} \approx 3.3 \times 10^{-2}$. On the other hand, our
theory predicts $\sigma_{_{B}} = (3 - 6) \times 10^{-2}$ depending
on the parameter $\mu$. Note that in the numerical simulations
\cite{B05,BHK05} the shear is not small (i.e., the parameter $S
\tau_0 \sim 1$), which explains some difference between the
theoretical predictions and numerical simulations. Therefore, the
numerical simulations \cite{B05,BHK05} clearly demonstrate the
existence of the large-scale dynamo in the absence of mean kinetic
helicity and alpha effect, in agreement with the theoretical
predictions discussed in the present paper.

\begin{acknowledgments}
We have benefited from stimulating discussions on the conditions for
the shear-current dynamo instability with Alexander Schekochihin. We
acknowledge valuable discussions with Valery Pipin. This work was
partially supported by the Royal Society. We thank for hospitality
the Department of Applied Mathematics and Theoretical Physics of
University of Cambridge.
\end{acknowledgments}

\appendix

\section{The electromotive force in a sheared turbulent convection}

\renewcommand{\theequation}
            {A.\arabic{equation}}

In order to study the shear-current effect in a sheared turbulent
convection we use a procedure applied in \cite{RK04,RK06} for
similar problems. Let us derive equations for the second moments. To
exclude the pressure term from the equation of motion~(\ref{B1}) we
calculate $\bec{\nabla} {\bf \times} (\bec{\nabla} {\bf \times} {\bf
u})$. Then we rewrite the obtained equation and
Eqs.~(\ref{B2})-(\ref{B3}) in a Fourier space. We also apply the
two-scale approach, e.g., we use large scale ${\bf R} = ( {\bf x} +
{\bf y}) / 2$, $\, {\bf K} = {\bf k}_1 + {\bf k}_2$ and small scale
${\bf r} = {\bf x} - {\bf y}$, $\, {\bf k} = ({\bf k}_1 - {\bf k}_2)
/ 2$ variables (see, e.g., \cite{RS75}). This implies that we assume
that there exists a separation of scales, i.e., the maximum scale of
turbulent motions $l_0$ is much smaller then the characteristic
scale $L_B$ of inhomogeneities of the mean magnetic field. We derive
equations for the following correlation functions:
\begin{eqnarray}
f_{ij}({\bf k}) &=& \hat L(u_i; u_j) \;, \quad h_{ij}({\bf k}) =
\hat L(b_i; b_j) \;,
\nonumber\\
g_{ij}({\bf k}) &=& \hat L(b_i; u_j) \;, \quad F_{i}({\bf k}) = \hat
L(s; u_i) \;,
\nonumber\\
G_{i}({\bf k}) &=& \hat L(s; b_i) \;, \quad \Theta({\bf k}) = \hat
L(s; s) \;,
\label{BBB1}
\end{eqnarray}
where
\begin{eqnarray*}
\hat L(a; c) = \int \langle a({\bf k} + {\bf  K} / 2) c(-{\bf k} +
{\bf  K} / 2) \rangle
\\
\times \exp{(i {\bf K} {\bf \cdot} {\bf R}) } \,d {\bf  K} \; .
\end{eqnarray*}
The equations for these correlation functions are given by
\begin{eqnarray}
{\partial f_{ij}({\bf k}) \over \partial t} &=& i({\bf k} {\bf
\cdot} {\bf B}) \Phi_{ij} + I^f_{ij} + I_{ijmn}^\sigma({\bf U})
f_{mn} + \hat{\cal N} f_{ij} \;,
\nonumber \\
\label{B6} \\
{\partial h_{ij}({\bf k}) \over \partial t} &=& - i({\bf k}{\bf
\cdot} {\bf B}) \Phi_{ij} + I^h_{ij} + E_{ijmn}^\sigma({\bf U})
h_{mn} + \hat{\cal N} h_{ij} \;,
\nonumber \\
\label{B7} \\
{\partial g_{ij}({\bf k }) \over \partial t} &=& i({\bf k} {\bf
\cdot} {\bf B}) [f_{ij}({\bf k}) - h_{ij}({\bf k}) - h_{ij}^{(H)}] +
I^g_{ij}
\nonumber \\
&& + J_{ijmn}^\sigma({\bf U}) g_{mn} + g e_n P_{jn}(k) G_{i}(-{\bf
k})
\nonumber \\
&& + \hat{\cal N} g_{ij} \;,
\label{B8} \\
{\partial F_{i}({\bf k}) \over \partial t} &=& - {\rm i}\,({\bf k}
{\bf \cdot} {\bf B}) G_{i}({\bf k})  + I^F_{i} + H_{im}^\sigma({\bf
U}) \, F_{m}
\nonumber \\
&& + g e_n P_{in}(k) \Theta({\bf k}) + \hat{\cal N} F_{i} \;,
\label{B9} \\
{\partial G_{i}({\bf k}) \over \partial t} &=& - {\rm i}\,({\bf k}
{\bf \cdot} {\bf B}) F_{i}({\bf k}) + I^G_{i} + (\nabla_m U_i) \,
G_{m}({\bf k})
\nonumber \\
&&+ \hat{\cal N} G_{i} \;,
\label{B10} \\
{\partial \Theta({\bf k}) \over \partial t} &=& - {N^2 \over g}
F_{z}({\bf k}) + \hat{\cal N} \Theta \;, \label{B11}
\end{eqnarray}
where hereafter we omit argument $t$ and ${\bf R}$ in the
correlation functions and neglect small terms $ \sim O(\nabla^2)$.
Here $\bec{\nabla} = \partial / \partial {\bf R} $, and we also
neglect a small term $\propto N^2 / g$ in Eq.~(\ref{B11}). In
Eqs.~(\ref{B6})-(\ref{B11}), $\Phi_{ij}({\bf k}) = g_{ij}({\bf k}) -
g_{ji}(-{\bf k})$, $\, P_{ij}(k) = \delta_{ij} - k_i k_j / k^2$, $\,
\hat{\cal N} f_{ij} = g e_n [P_{in}(k) F_{j}({\bf k}) + P_{jn}(k)
F_{i}(-{\bf k})] + \hat{\cal N} \tilde f_{ij}$, and $ \hat{\cal
N}\tilde f_{ij}$, $\, \hat{\cal N}h_{ij}$, $\, \hat{\cal N}g_{ij}$,
$\, \hat{\cal N}F_{i}$, $\, \hat{\cal N}G_{i}$ and $\hat{\cal
N}\Theta$ are the third-order moment terms appearing due to the
nonlinear terms. The terms which are proportional to the heat flux
$F_i$ in the tensor $ \hat{\cal N}f_{ij}$, can be considered as a
stirring force for the turbulent convection. Note that a stirring
force in the Navier-Stokes turbulence is an external parameter. The
tensors $I_{ijmn}^\sigma({\bf U})$, $\, E_{ijmn}^\sigma({\bf U})$,
$\, J_{ijmn}^\sigma({\bf U})$ and $H_{ij}^\sigma({\bf U})$ are given
by
\begin{eqnarray*}
I_{ijmn}^\sigma({\bf U}) &=& \biggl[2 k_{iq} \delta_{mp} \delta_{jn}
+ 2 k_{jq} \delta_{im} \delta_{pn} - \delta_{im} \delta_{jq}
\delta_{pn}
\nonumber\\
&& - \delta_{iq} \delta_{jn} \delta_{pm} + \delta_{im} \delta_{jn}
k_{q} {\partial \over \partial k_{p}} \biggr] \nabla_{p} U_{q} \;,
\nonumber\\
E_{ijmn}^\sigma({\bf U}) &=& \biggl[\delta_{im} \delta_{jq}
\delta_{pn} + \delta_{jm} \delta_{iq} \delta_{pn}
\nonumber\\
& & + \delta_{im} \delta_{jn} k_{q} {\partial \over \partial k_{p}}
\biggr] \nabla_{p} U_{q} \;,
\nonumber\\
J_{ijmn}^\sigma({\bf U}) &=& \biggl[2 k_{jq} \delta_{im} \delta_{pn}
- \delta_{im} \delta_{pn} \delta_{jq} + \delta_{jn} \delta_{pm}
\delta_{iq}
\nonumber\\
& & + \delta_{im} \delta_{jn} k_{q} {\partial \over \partial k_{p}}
\biggr] \nabla_{p} U_{q} \;,
\nonumber\\
H_{ij}^\sigma({\bf U}) &=& 2 k_{in} \nabla_{j} U_{n} - \nabla_{j}
U_{i} \;,
\end{eqnarray*}
where $ k_{ij} = k_i k_j / k^2 $. The source terms $I_{ij}^f$ , $\,
I_{ij}^h$, $\, I_{ij}^g$, $\,I^F_{i}$ and $\,I^G_{i}$ which contain
the large-scale spatial derivatives of the mean magnetic field, are
given in \cite{RK04,RK06} (see also Eqs.~(\ref{M2}) and~(\ref{M3})
below). Next, in Eqs.~(\ref{B6})-(\ref{B11}) we split the tensor for
magnetic fluctuations into nonhelical, $h_{ij},$ and helical,
$h_{ij}^{(H)},$ parts. The helical part of the tensor of magnetic
fluctuations $h_{ij}^{(H)}$ depends on the magnetic helicity and it
follows from the magnetic helicity conservation arguments (see, {\rm
e.g.,} \cite{KR82,KRR94,GD94,KR99,KMRS2000,BB02}). We also use the
spectral $\tau$ approximation which postulates that the deviations
of the third-moment terms, $\hat{\cal N} M^{(III)}({\bf k})$, from
the contributions to these terms afforded by the background
turbulent convection, $\hat{\cal N} M^{(III,0)}({\bf k})$, are
expressed through the similar deviations of the second moments,
$M^{(II)}({\bf k}) - M^{(II,0)}({\bf k})$ [see Eq.~(\ref{AAC3})].

We take into account that the characteristic time of variation of
the mean magnetic field ${\bf B}$ is substantially larger than the
correlation time $\tau(k)$ for all turbulence scales. This allows us
to get a stationary solution for Eqs.~(\ref{B6})-(\ref{B11}) for the
second-order moments, $M^{(II)}({\bf k})$, which are the sum of
contributions caused by a shear-free turbulent convection and a
sheared turbulent convection. The contributions to the mean
electromotive force caused by a shear-free turbulent convection are
given in \cite{RK06}. On the other hand, the contributions to the
mean electromotive force caused by the sheared turbulent convection
are ${\cal E}_{m}^{\sigma} = \varepsilon_{mji} \, \int \,
g_{ij}^{\sigma}({\bf k}) \,d {\bf k} $. In particular, in the
kinematic approximation the contributions to the cross-helicity
tensor $g_{ij}^{\sigma}$ caused by the sheared turbulent convection,
are given by
\begin{eqnarray}
g_{ij}^{\sigma}({\bf k}) &=& \tau [J_{ijmn}^\sigma \tilde g_{mn} +
I^{(g,\sigma)}_{ij} + g e_n P_{jn}(k) G_{i}^\sigma(-{\bf k})] \;,
\nonumber\\
\label{S5}
\end{eqnarray}
where
\begin{eqnarray}
G_{i}^\sigma({\bf k})  &=& \tau^2 \, (\nabla_m U_i) \, I^G_{m}  \;,
\label{S6}\\
\tilde g_{ij} &=& \tau \, [I_{ij}^g + \tau \, g \, e_n \, P_{jn}(k)
\, I^G_{i}] \;,
\label{M4}\\
I^{(g,\sigma)}_{ij} &=& \tau \, [(2 \, P_{js}(k) - \delta_{js}) \,
E_{ikmn}^\sigma \, h_{mn}^{(0)}
\nonumber\\
& & -   \delta_{is} \, I_{kjmn}^\sigma \, f_{mn}^{(0)}] \,B_{s,k}
\;,
\label{M1}\\
I^G_{i} &=& - \biggl[\delta_{ij} \, \delta_{mk} + {1 \over 2} \,
\delta_{im} \, k_{j} \, {\partial \over \partial k_{k}}\biggr] \,
F_{m}^{(0)} \, B_{j,k}  \;,
\nonumber\\
\label{M2}\\
I_{ij}^g &=& \biggl[(2 \, P_{jn}(k) - \delta_{jn}) \, h_{ik}^{(0)} -
\delta_{in} \, f_{kj}^{(0)}
\nonumber\\
& & - {1 \over 2} \, k_{n} \, {\partial \over \partial k_{k}}
(f_{ij}^{(0)} + h_{ij}^{(0)})\biggr] \, B_{n,k} \;, \label{M3}
\end{eqnarray}
and $B_{i,j} = \nabla_j B_{i}$. We take into account that in
Eq.~(\ref{S5}) the terms with symmetric tensors with respect to the
indexes "i" and "j" do not contribute to the mean electromotive
force because ${\cal E}_{m}^{\sigma} = \varepsilon_{mji} \, \int \,
g_{ij}^{\sigma}({\bf k}) \,d {\bf k} $. For the integration in ${\bf
k}$-space we have to specify a model for the background shear-free
turbulent convection (with ${\bf B} = 0)$, which is determined by
Eqs.~(\ref{B15})-(\ref{B17}) in Section III.

The contributions to the mean electromotive force caused by the
sheared turbulent convection, are ${\cal E}_{i}^{\sigma} =
b_{ijk}^{\sigma} \, \nabla_k \, B_{j}$, where the tensor
$b_{ijk}^{\sigma} = b_{ijk}^{u} + b_{ijk}^{F}$ is given by
Eqs.~(\ref{B18})-(\ref{B19}) in Section III. For derivation of
Eqs.~(\ref{B18})-(\ref{B19}) we use the following identities:
\begin{eqnarray*}
\int {k_{i} k_{j} k_{m} k_{n} \over k^{4}} \sin \theta \,d \theta
\,d \varphi &=& {4 \pi \over 15} \, \triangle_{ijmn} \;,
\\
\int {k_{i} k_{j} k_{m} k_{n} k_{p} k_{q} \over k^{6}} \sin \theta
\,d \theta \,d \varphi &=& {4 \pi \over 105} \, \triangle_{ijmnpq}
\;,
\end{eqnarray*}
and
\begin{eqnarray*}
&& \triangle_{ijmn} = \delta_{ij} \, \delta_{mn} + \delta_{im} \,
\delta_{nj} + \delta_{in} \, \delta_{mj} \;,
\\
&& \triangle_{ijmnpq} = \triangle_{mnpq} \, \delta_{ij} +
\triangle_{jmnq} \, \delta_{ip} + \triangle_{imnq} \, \delta_{jp}
\\
&& \; + \triangle_{jmnp} \, \delta_{iq} + \triangle_{imnp} \,
\delta_{jq} + \triangle_{ijmn} \, \delta_{pq} - \triangle_{ijpq} \,
\delta_{mn} \;,
\end{eqnarray*}
and
\begin{eqnarray*}
&& \varepsilon_{ikm} \, e_{ns} \, \triangle_{jpqmns} \, \nabla_p \,
U_{q} = 2 \, \varepsilon_{ikm} \, [(\partial U)_{mj}
\\
&& \quad + 2 \, e_{mp} \, \nabla_p \, U_{j} + 2 \, e_{qj} \,
\nabla_m \, U_{q}] \;,
\\
&& \varepsilon_{inm} \, e_{ms} \, \triangle_{jkpqns} \, \nabla_p \,
U_{q} = 4 \, \{\varepsilon_{inm} \, [e_{mk} \, (\partial U)_{nj}
\\
&& \quad + e_{mj} \, (\partial U)_{kn}] + e_{mn} \,
[\varepsilon_{ikm} \, (\partial U)_{nj} + \varepsilon_{ijm} \,
(\partial U)_{kn}] \} \;,
\\
&& \varepsilon_{inq} \, e_{ms} \, \triangle_{jkpmns} \, \nabla_p \,
U_{q} = \varepsilon_{inq} \, \{\nabla_p \, U_{q} \, [2 \, e_{mk} \,
\triangle_{jpmn}
\\
&& \quad + 2 \, e_{mj} \, \triangle_{kpmn} + \triangle_{jpkn}] -
\nabla_n \, U_{q} \, (\delta_{jk} + 2 e_{jk}) \} \;,
\\
&& \varepsilon_{ijn} \, e_{ms} \, \triangle_{kpqmns} \, \nabla_p \,
U_{q} = 2 \, \{\varepsilon_{ijn} \, [2\, e_{mk} \, (\partial U)_{nm}
\\
&& \quad + 2\, e_{mn} \, (\partial U)_{km} +  (\partial U)_{kn}] +
\varepsilon_{ijk} \, e_{mn} \, (\partial U)_{nm} \} \; .
\end{eqnarray*}
In Eqs.~(\ref{B18})-(\ref{B19}) we have taken into account only the
terms which contribute to the shear-current effect. In particular,
we consider the mean magnetic field in a most simple form ${\bf B} =
(B_x(z), B_y(z), 0)$ and we take into account that $B_y \gg B_x$ and
$S \, \tau_0 \ll 1$, where the mean velocity shear is ${\bf U} = (0,
Sx, 0)$ and $ {\bf W} = (0,0,S)$. Straightforward calculations using
Eqs.~(\ref{B18})-(\ref{B19}) and Eqs.~(\ref{SB20}) and~(\ref{SB21})
yield $\sigma_{_{B}} = \sigma^u_{_{B}} + \sigma^F_{_{B}}$, where
\begin{eqnarray}
\sigma^u_{_{B}} &=& {(q-1+\mu)^2 \over 15 \, (q-1+2 \mu ) \, (q-1)}
\, [1 - \mu + \epsilon \, (1 + 2 \, \mu)] \;,
\nonumber\\
\label{B20}\\
\sigma^F_{_{B}} &=& {a_\ast \, (q-1+\mu)^3 \over 35 \, (q-1+3 \mu )
\, (q-1)^2} \, [2 + 3 \, (2 - 3 \mu ) \, \sin^2 \, \phi ] \;,
\nonumber\\
\label{B21}
\end{eqnarray}
$\phi$ is the angle between the unit vector ${\bf e}$ and the
background vorticity ${\bf W}$ due to the large-scale shear.
Equations~(\ref{B20})-(\ref{B21}) yield Eq.~(\ref{E1}) given in
Sect. III.

\end{document}